# From Instructor to Collaborator: What a 90-Participant Study Reveals about Human–Agent Collaboration in a Mobile Serious Game

Danai Korre

University of Bedfordshire, danai.korre@beds.ac.uk

This position paper reflects empirical data collected during my PhD from a large-scale within-subjects study (N = 90). The study compared a highly human-like, spoken embodied conversational agent (ECA) against a low human-like text base agent (no embodiment, text bubble only) within a mobile, Unity-developed game about pre-decimal UK currency. The game included two agents with different roles-an Instructor (Alex) and a Shopkeeper/Collaborator. Users interacted using voice and mouse input. The quantitative data I collected included a usability questionnaire (CCIR MINERVA) and the Agent Persona Instrument. Data was analyzed using paired t-test, repeated measures ANOVA and multiple linear regression to identify correlations between the persona and usability. The results showed a statistically significant preference for the version of highly human-like agents, with a large effect size. This is further discussed alongside qualitative findings from observations and exit interviews. The results are framed for Human-Agent collaboration, especially for how roles, mixed-initiative dialogue, and breakdowns/repairs become apparent in goal-oriented tasks. I conclude with questions on timing, user expectations, and role-specific interactions. This submission does not propose new frameworks; it reports empirical findings and questions I hope to workshop with the community.

## 1 MOTIVATION & BACKGROUND

The aim of this research is to examine how spoken humanoid embodied conversational agents (HECAs) can foster usability in a mobile serious game. Previous research on ECAs has shown that visual and behavioral human-likeness can influence perceived competence and social expectations (Cassell et al., 2000). This research also tries to answer how humanizing ECAs affect usability while contributing empirically to the area of conversational agents. This aligns with longstanding evidence that humans involuntarily attribute social agency to computers (Reeves & Nass, 1996).

At the time of the study, there was a growing body of empirical data on the effects of ECAs, but there were few empirical evaluations suggesting that ECAs are more usable on mobile devices or in serious games. Lack of evidence on the potential effects of ECAs within serious games hinders their utilization in serious games, as their design and development are a time-consuming and resource-intensive process without guaranteed benefits. The constructs used for this study were usability (CCIR MINERVA) (CCIR, 2004), agent persona (API) (Baylor & Ryu, 2003), and the "illusion of humanness" (users' involuntary mental response that the interface possesses human attributes and/or cognitive functions).

## 2 STUDY CONTEXT: MONEYWORLD, ROLES, AND MODALITIES

The application is a Unity-based, speech-enabled (PocketSphinx) mobile serious game called Moneyworld. The participants “time-travel” to a 1960s corner shop to perform coin-based purchases of everyday items thus understanding inflation and the old monetary system.

The game includes two agents, the instructor (“Alex”) for tutorial/background [Figure 1(a) highly human-like 1(c) low human-like] and the Shopkeeper/Collaborator for transactional dialogue and validation [Figure 1(c)].

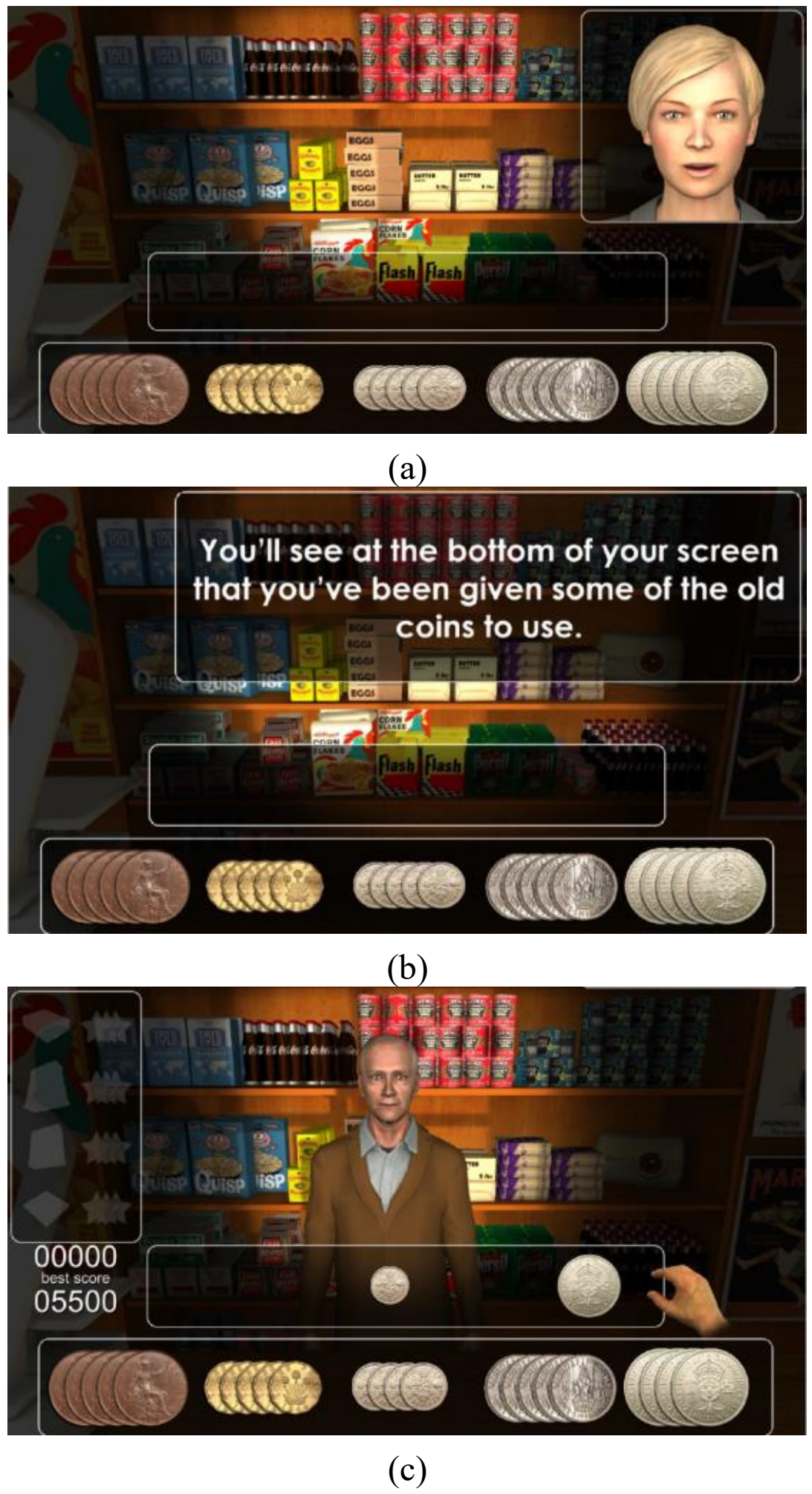

Figure 1: Two conversational agents support the task: (a,b) an Instructor agent (“Alex”) for tutorial guidance and (c) a Shopkeeper/Collaborator for transactional dialogue. Users interact through spoken dialogue and mouse-based coin selection. (Adapted from Korre, 2019)

Research on pedagogical agents has shown that clear instructional versus collaborative roles shape user expectations and perceived social presence (Baylor, 2011). The participants used multimodal interaction, speech for dialogue, and a mouse for coin selection/submission.

## 3 METHOD

The study used a within-subjects (repeated-measures) 2x2 factorial design, balanced for order to avoid ordering bias. Each participant experienced both versions, followed by questionnaires and an exit interview.

The experiment planned N=90, following power guidance for regression (10–15 samples per 9 predictors). Participants were divided into balanced groups. The metrics used included one for usability-CCIR MINERVA (18 items; 7-pt), and one for agent persona-API (24 items; 5-pt; factors: facilitating learning, credible, human-like, engaging) (API). Both are standardized and validated metrics. Analysis was conducted using paired t-tests, repeated-measures ANOVA, and multiple linear regression (affective-persona predictors - usability DV) with assumption checks.

## 4 QUANTITATIVE RESULTS

The main outcome is that users prefer interacting with HECAs. Difference between the two versions is statistically significant for all conditions with a large effect size (d = 1.01). Figure 2 presents the mean usability scores for the different versions as well as the mean API scores for the different agents.

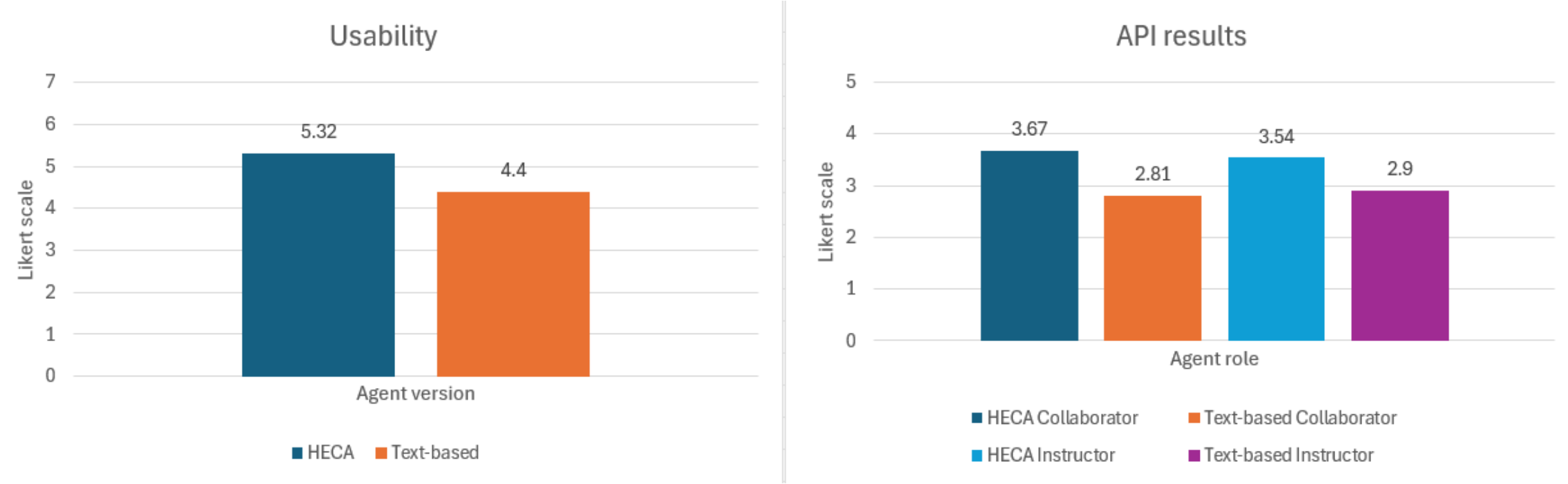

Figure 2: Summary of reported differences. The HECA condition was rated higher in overall usability and agent persona than the text-agent condition. See thesis for exact statistics and effect sizes (Korre, 2019, Ch. 5).

The affective API persona items (e.g., human-like, expressive, friendly, entertaining, motivating) acted as predictors of usability for both the shopkeeper and the instructor agents.

## 5 QUALITATIVE INSIGHTS

Participants described the spoken, animated HECA as more engaging and natural in both instructional and collaborative/transactional interactions. Breakdowns (e.g., speech recognition) required concise repairs during the task, emphasizing timing/placement of collaborative support.

## 6 DISCUSSION FOR W30: COLLABORATION LENS ON THE FINDINGS

Findings highlight the way agent roles affect human-agent collaboration. Participants had different expectations from the Shopkeeper/Collaborator and Instructor agents regarding guidance and initiative. They have also relied on the agent to coordinate mixed-initiative interaction, which aligns with mixed-initiative principles in human-AI interaction (Amershi et al., 2019). Speech breakdowns needed solid repairs to achieve common ground, while speech recognition issues disrupted task flow. Finally, we used human voice recordings for the HECA agents, which resulted in high human-like voices, which, along with the agent's embodiment, increased user expectations in terms of competence. This underscores the need to align capability, role, and appearance. Together, these insights suggest opportunities for clearer role signaling, better grounding support, and more deliberate expectation-management in collaborative agents.

## 7 LIMITATIONS

The study must be approached with caution regarding generalizability, as it was conducted within the specific context of a mobile serious game. Also, speech recognition constrains formed patterns of breakdown and repair that may not be observed under more controlled environments. Finally, the study focused on usability and agent persona. More sensitive collaboration metrics, learning outcomes, and long-term effects were outside of the scope of the study.

## 8 QUESTIONS TO DISCUSS AT W30

- What behaviors constitute "role-appropriate" interaction for Instructor versus Collaborator agents, and how should these be evaluated?
- How can designers calibrate user expectations when highly human-like embodiment raises assumptions about competence and initiative?
- What lightweight metrics could capture collaboration quality (e.g., initiative balance, repair efficiency) in short, in-the-wild studies?
- When multiple agents with different roles are present, what handover cues or artefacts best maintain user understanding across transitions?
- How should collaboration cues be timed to support smooth progress without overwhelming or interrupting the user?

## ACKNOWLEDGMENTS

This PhD research was conducted at the University of Edinburgh; I thank my supervisors, Professors Austin Tate and Judy Robertson for their useful feedback and constructive recommendations, funders and collaborators who are acknowledged in the thesis.